\documentclass[prd,twocolumn,showpacs,preprintnumbers,amsmath,amssymb]{revtex4}
\usepackage{graphics}
\usepackage{epsfig}
\usepackage{dcolumn}% Align table columns on decimal point
\usepackage{bm}% bold math

\def\beq{\begin{eqnarray}}

\def\eeq{\end{eqnarray}}
\def\non{\nonumber}

\def\non{\nonumber}

\def\la{\langle}
\def\ra{\rangle}

\begin{document}
%\begin{CJK}{GBK}{song}
\title{Contribution of final state interaction to the branching ratio of $B\to J/\psi D$}

%\author{Xiang Liu£¨ÁõÏ裩$^{1}$}\email{liuxiang@teor.fis.uc.pt}\author{Zheng-Tao
%Wei£¨ÎºÕýÌΣ©$^2$}\email{weizt@nankai.edu.cn}
%\author{Xue-Qian Li£¨ÀîѧDZ£©$^2$}\email{lixq@nankai.edu.cn}

\author{Xiang Liu$^{1}$}\email{liuxiang@teor.fis.uc.pt}\author{Zheng-Tao
Wei$^2$}\email{weizt@nankai.edu.cn}
\author{Xue-Qian Li$^2$}\email{lixq@nankai.edu.cn} \affiliation{$^1$Centro de F\'{i}sica
Te\'{o}rica, Departamento de F\'{i}sica, Universidade de Coimbra,
P-3004-516 Coimbra,
Portugal\\
$^2$Department of Physics, Nankai University, Tianjin 300071,
China}

\date{\today}% It is always \today, today,

\begin{abstract}

To testify the validity of the perturbative QCD (pQCD) and
investigate its application range, one should look for a suitable
process to do the job. $B\to J/\psi D$ is a promising candidate.
The linear momentum of the products is relatively small, so that
there may exist a region where exchanged gluons are soft and the
perturbative treatment may fail, so that the non-perturbative
effect would be significant. We attribute such non-perturbative
QCD effects into the long-distance final state interaction (FSI)
which is estimated in this work. We find that the contribution
from the FSI to the branching ratio is indeed sizable and may span
a rather wide range of $10^{-6}\sim 10^{-5}$, and cover a region
where the pQCD prediction has the same order. A more accurate
measurement on its branching ratio may provide important
information about the application region of pQCD and help to
clarify the picture of the inelastic rescattering (i.e. FSI) which
is generally believed to play an important role in $B$ decays.

\end{abstract}


\maketitle

%%%%%%%%%%%%%%%%%%%%%%%%%%%%%%%%%%%%
\section{introduction}\label{sec1}
%%%%%%%%%%%%%%%%%%%%%%%%%%%%%%%%%%%%

It is well known that $B$ physics may provide an ideal field for
testing all the existing theoretical frameworks, methods and
searching new physics beyond the standard model (SM). The reason
is that because there exists a heavy flavor, some approximations,
such as the $1/M_Q$ expansion can be adopted,  so that the results
of perturbative calculations are more reliable. When the study
gets deeper, defects of such theoretical frameworks have been
unavoidably exposed and demand further improving the standing
theories. The most fundamental problem is how to properly evaluate
the hadronic matrix elements which are fully governed by
non-perturbative QCD. Thanks to the factorization, one can
separate the non-perturbative QCD effects from the perturbative
parts and the later one is calculable in terms of the field theory
order by order. Based on this picture, various theories, such as
the naive QCD factorization, pQCD (perturbative QCD) and the
soft-collinear-effective theory (SCET) etc. are invented to
calculate the processes where $B$ mesons are involved.

The pQCD has been proved to be a successful approach in $B$
physics, namely most of the results obtained in this approach are
consistent with data of the Babar, Belle and CLEO experiments. In
this approach, the infrared divergence is properly dealt with by
taking into account the contribution of the transverse momentum of
quarks $k_T$. In this picture, the non-perturbative part is
included in the wavefunctions of the initial $B$ meson and the
produced hadrons. Obviously, as one factorizes the perturbative
part out and calculates the quark-level transition amplitudes, he
must assume that all the constituents which participate in the
reaction are not far from their mass shells and moreover, all the
internal lines (no matter quark line or gluon line) must be hard
enough, so that the perturbative calculation can make sense. A
natural question would be raised, whether the pQCD framework is
complete, even though its validity is supposed to be respected. By
the asymptotic freedom of QCD at higher energy region, the
perturbative approach works perfectly well, however, if one or two
internal lines can reach a low-energy region where they are not
sufficiently hard, one can be convinced that at this region the
pQCD approach fails, or cannot result in reasonable values. If the
internal lines are soft, one can conjecture that at this region
the non-perturbative QCD would dominate and it could be attributed
into the so-called FSI, or the re-scattering sub-processes. To
identify the application range of pQCD and testify its validity,
we need to look for such processes where some internal lines can
be soft.

The process $B\to J/\psi+D$ just serves for the purpose. The
direct weak transition of $B\to J/\psi+D$ occurs via annihilation
between $b$ and $\bar u$ or $W$-exchange between $b$ and $\bar d$
(usually, we just name both of them as "annihilation"), and a pair
of $c\bar c$ can emerge from a gluon splitting. Since the $m_B\sim
5.3$ GeV, $m_{J/\psi}\sim 3.1$ GeV, and $m_D\sim 1.87$ GeV, and
the linear momentum of the products in the CM frame of $B$ meson
is as small as 0.9 GeV, thus there exists a region where the
gluon-line is soft. Thus one needs to include the contribution
from the long-distance effects in the theoretical calculations as
well as the short-distance effects which are contributed by the
hard gluon lines. The strategy is following. There have been some
calculations on the decay width for $B\to J/\psi+D$ in terms of
pQCD approach which we suppose to be the contribution of the
direct transition from $B$ into the final state $J/\psi+D$, and
then in this work, we calculate the contribution from
long-distance effects to the rates, and then we urge our
experimental colleagues to carry out an more accurate measurement
to testify the validity of the whole theoretical framework.

There have been some experimental attempts along the line. The
CLEO collaboration once reported a slow $J/\psi$ bump in the
inclusive spectrum of $B \to J/\psi+X$ \cite{Cleo-bump}, which
later was confirmed by Belle \cite{Belle-bump} and Babar
\cite{Babar-bump}. These experiments indicate that there exists an
excess in the momentum spectrum of the $J/\psi$ recoiling mass at
about $2$ GeV. The branching ratio of the excess is $6\times
10^{-4}$. Along with these experiments, different theoretical
explanations have been suggested in Ref. \cite{brod,hou,EMY}.

These theoretical works were focusing on the calculations of
direct transitions. Accompanying these theoretical hypotheses
\cite{brod,hou,EMY}, theorist and experimentalist indeed began to
study the branching ratio of $B\to J/\psi D$. Assuming the
intrinsic charm $c\bar c$ inside the $B$ meson, Chang and Hou
suggested that the branching ratio of $B\to J/\psi$ should reach
an order of magnitude of $10^{-4}$ \cite{hou}. In 2002, by the
collinear factorization approach, Eilam, Ladisa and Yang once
calculated the branching ratio of inclusive $B\to J/\psi$, and
obtained it to be $7.28\times 10^{-8}$ \cite{EMY}. The Babar and
Belle collaborations reported a negative result for searching
$B\to J/\psi \bar D$ decay. The upper limits on the branching
fractions is set as $1.3 \times 10^{-5}$ and $2.0 \times 10^{-5}$
for $B^0\to J/\psi \bar D^0$ respectively corresponding to the
Babar and Belle experiments \cite{Babar,Belle}, which show that
the assumption of the intrinsic charm $c\bar c$ inside the $B$
meson should be excluded. Later Li, Lu and Qiao reexamined the
$B\to J/\psi$ in the framework of the pQCD $k_T$ factorization,
and predicted $B[B^0\to J/\psi D]=3.45\times 10^{-6}$ \cite{LLQ}.

As discussed above, besides the theoretical calculations on the
decay width in terms of pQCD, one needs to take the FSI more
seriously. In this work, we are going to evaluate this
contribution in terms of the hadronic loops
\cite{Liu,Hadronloop-1,Hadronloop-2,HY-Chen}. Namely, we consider
several sub-processes such as $B\to D^{(*)}\pi (D^{(*)}\rho) \to
J/\psi D$. Here we suppose the transition hamiltonian can be
written as a sum
$$H=\sum_i^j H_i,$$
where $H_i$ corresponds to both the quark-level and hadron-level
hamiltonians and then
\begin{eqnarray*}
\langle J/\psi D|H|B\rangle&=&\langle J/\psi
D|H_{quark}|B\rangle\nonumber\\&&+\langle J/\psi
D|H_{had}|n\rangle \langle n|H'_{quark}|B\rangle+...
\end{eqnarray*}
where $H_{quark}$ and $H'_{quark}$ are the hamiltonian at quark
level, but contribute to different states (for example $J/\psi D$
or $D^{(*)}\pi,\; D^{(*)}\rho$ etc.) and the intermediate states
$|n\rangle$ are  the corresponding states with appropriate quantum
numbers. Indeed $\langle J/\psi D|H_{had}|n\rangle$ is just the
inelastic re-scattering amplitude which should be evaluated.

The above formulation indicates that the two parts should
interfere, but the relative phase between the two parts (or
several parts) is hard to determine because the different
amplitudes are caused by different hamiltonians and there (so far)
is no any symmetry to associate them yet. To estimate the order of
magnitude of such long-distance effects, we simply suppose the
interferences among different modes are constructive. The first
matrix element $\langle n|H'_{quark}|B\rangle$ where $|n\rangle$
can be $D\pi$ etc., can be evaluated reliably in terms of pQCD,
since there is a sufficient phase space.

Based on the idea, we re-evaluate the branching ratio of $B\to
J/\psi D$ by considering the contributions from the hadronic loop
effect to $B^0\to J/\psi \bar D^0$.

This paper is organized as follow. We present the calculation of
Hadronic loop contribution for $B^0\to J/\psi\bar D^0$ in Sec.
\ref{sec3}. Then we present the formulation about the
factorization of $B\to D^{(*)}\pi(\rho)$ in Sec. \ref{sec2}. In
Sec. \ref{sec4}, the numerical result is given. The last section
is a short conclusion and discussion.

%%%%%%%%%%%%%%%%%%%%%%%%%%%%%%%%%%%%%%%%%%%%%%%%%%%%%%%%%%%%%%%%%
\section{Hadronic loop contribution for $B^0\to J/\psi\bar D^0$}\label{sec3}
%%%%%%%%%%%%%%%%%%%%%%%%%%%%%%%%%%%%%%%%%%%%%%%%%%%%%%%%%%%%%%%%%

\begin{figure}[htb]
\begin{center}
\begin{tabular}{ccccccc}
\scalebox{0.4}{\includegraphics{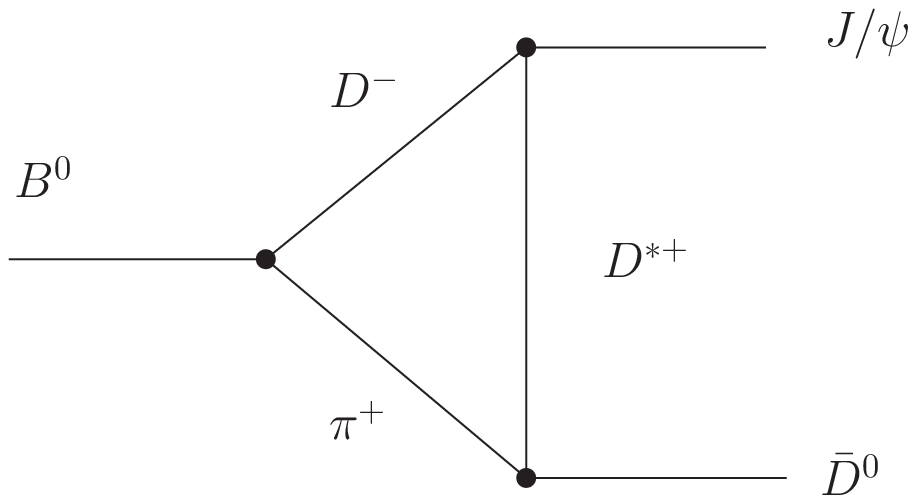}}&\scalebox{0.4}{\includegraphics{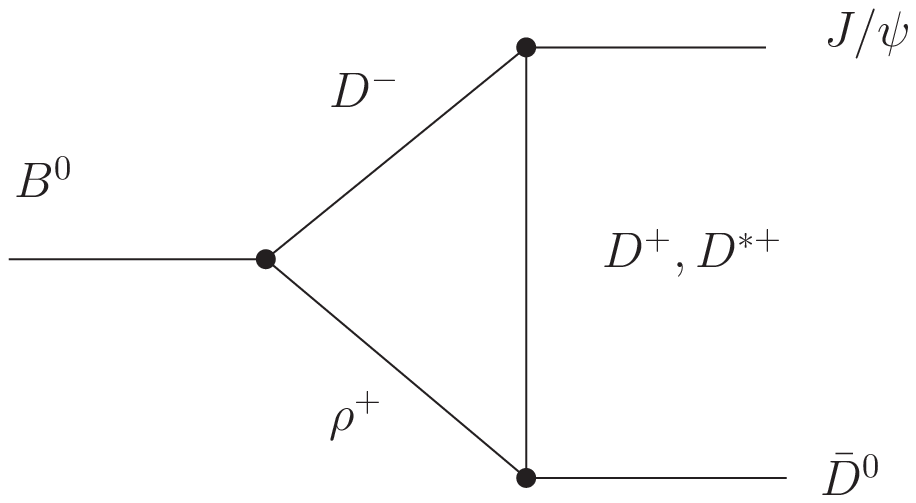}}\\\\
(a)&(b)\\\\
\scalebox{0.4}{\includegraphics{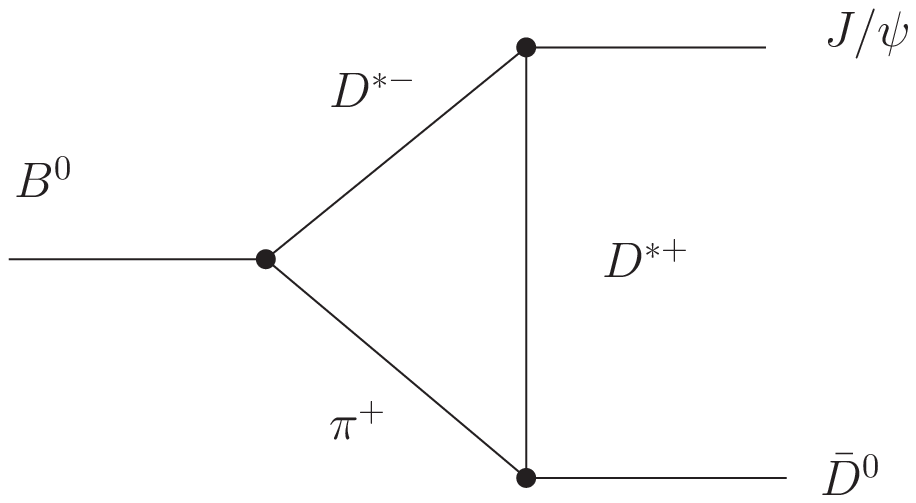}}&\scalebox{0.4}{\includegraphics{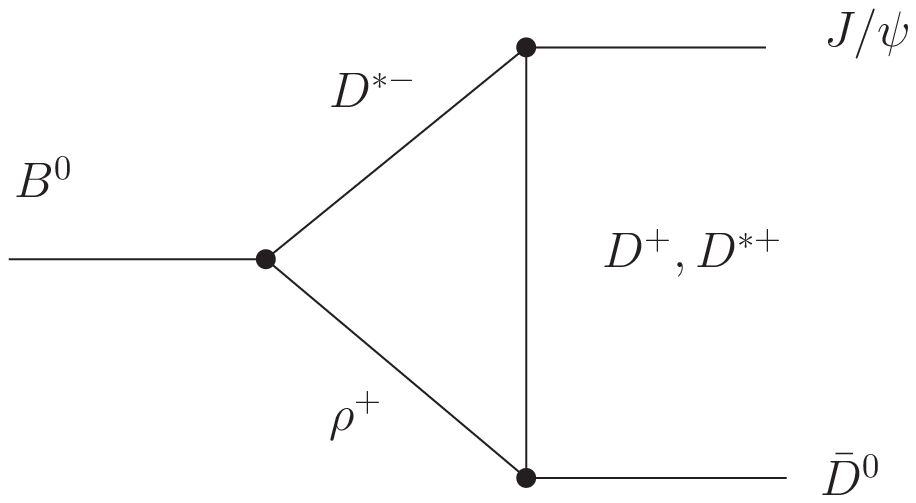}}\\\\
(c)&(d)\\\\
\scalebox{0.4}{\includegraphics{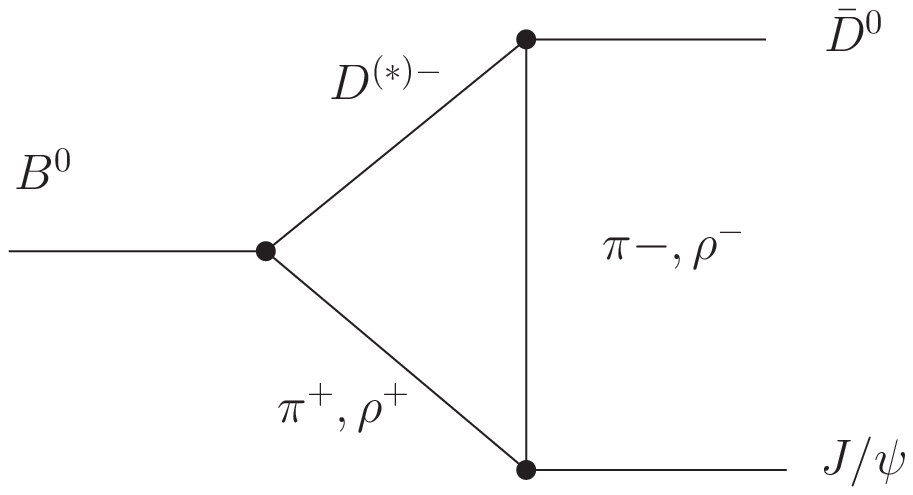}}&\\\\
(e)&\\
\end{tabular}
\end{center}
\caption{The hadronic loop diagrams depict the hadronic loop
effect on $B^0\to J/\psi\bar{D}^0$.\label{diagrams}}
\end{figure}

\begin{figure}[htb]
\begin{center}
\begin{tabular}{c}
\scalebox{0.45}{\includegraphics{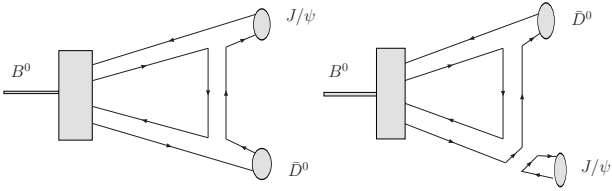}}\\
\end{tabular}
\end{center}
\caption{Left-hand and right-hand sides diagrams respectively
correspond to the OZI allowed and OZI forbidden diagrams for the
hadronic loop diagrams for decay of $B^0\to J/\psi\bar{D}^0$.
\label{compariation}}
\end{figure}

The diagrams which determine the hadronic loop effects on the rate
of $B^0\to J/\psi \bar D^0$ decay are depicted in Fig.
\ref{diagrams}, which can be divided into two groups. The fist group
includes Fig. \ref{diagrams} (a)-(d) and exactly corresponds to the
left diagram of Fig. \ref{compariation} which is depicted by a
process at the quark level. Definitely, there are quark lines
flowing  from initial state hadron to the final state ones and
therefore are the OZI allowed. Another group including only Fig.
\ref{diagrams} (e) corresponds to the the quark-level process and is
shown at the right diagram of Fig. \ref{compariation} and obviously
is an OZI forbidden diagram.

According to the OZI rule, the contribution of Fig. \ref{diagrams}
(e) is much suppressed comparing with that from Fig. \ref{diagrams}
(a)-(d). This fact can be confirmed by comparing the coupling
constants of $D^{(*)}D^{(*)}J/\psi$ and $\rho\pi J/\psi$.
$g_{\rho\pi J/\psi}$ is about three orders larger than that of
$g_{D^{(*)}D^{(*)}J/\psi}$ \cite{Liu}. Thus we can safely ignore the
contribution from Fig. \ref{diagrams} (e).

The effective Lagrangians at the hadron-hadron vertices in Ref.
\cite{Lagarangian} are
\begin{eqnarray}
\mathcal{L}_{\psi DD} &=& i g_{\psi DD}^{} \psi_\mu \left(
\partial^\mu D \bar{D} - D \partial^\mu \bar{D} \right),
\\
{\mathcal{L}}_{\psi D^* D^*} &=& -i g_{\psi D^* D^*}^{} \Bigl\{
\psi^\mu \left( \partial_\mu D^{*\nu} \bar{D}_\nu^* - D^{*\nu}
\partial_\mu \bar{D}_\nu^* \right)\nonumber \\ &&
+ \left( \partial_\mu \psi_\nu D^{*\nu} - \psi_\nu \partial_\mu
D^{*\nu} \right) \bar{D}^{*\mu} \nonumber \\ && + D^{*\mu} \left(
\psi^\nu \partial_\mu \bar{D}^*_\nu - \partial_\mu \psi_\nu
\bar{D}^{*\nu} \right) \Bigr\},
\\
{\mathcal{L}}_{\psi D^* D} &=& -g_{\psi D^* D}^{}
\varepsilon^{\mu\nu\alpha\beta} \partial_\mu \psi_\nu \left(
\partial_\alpha D^*_\beta \bar{D} + D \partial_\alpha \bar{D}^*_\beta
\right),\\
{\mathcal{L}}_{D^* D \pi} &=& i g_{D^*D\pi}^{} \left( D_\mu^*
\partial^\mu \pi \bar{D} - D \partial^\mu \pi \bar{D}^*_\mu
\right),
\\
{\mathcal{L}}_{D^* D^* \pi} &=& -g_{D^* D^* \pi}^{}
\varepsilon^{\mu\nu\alpha\beta} \partial_\mu D^*_\nu \pi
\partial_\alpha \bar{D}^*_\beta,
\\
{\mathcal{L}}_{DD\rho} &=& i g_{DD\rho}^{} \left( D \rho^\mu
\partial_\mu \bar{D} - \partial_\mu D \rho^\mu \bar{D} \right),
\\
%{\mathcal{L}}_{D^*D^*\rho} &=& i g_{D^*D^*\rho}^{}  \Bigl\{
%\partial_\mu D^*_\nu \rho^\mu \bar{D}^{*\nu} - D_\nu^* \rho_\mu
%\partial^\mu \bar{D}^{*\nu}
%+ \left( D^{*\nu} \partial_\mu \rho_\nu - \partial_\mu D_\nu^*
%\rho^\nu \right) \bar{D}^{*\mu} \nonumber \\ && \mbox{}
%\qquad\quad + D^{*\mu} \left( \rho^\nu \partial_\mu \bar{D}^*_\nu
%- \partial_\mu \rho_\nu \bar{D}^{*\nu} \right) \Bigr\},
%\\
{\mathcal{L}}_{D^* D \rho} &=& -g_{D^* D \rho}^{}
\varepsilon^{\mu\nu\alpha\beta} \left( D \partial_\mu \rho_\nu
\partial_\alpha \bar{D}^*_\beta + \partial_\mu D^*_\nu \partial_\alpha
\rho_\beta \bar{D} \right).\nonumber\\
\end{eqnarray}
Here $\varepsilon_{0123} = +1$, $\pi =\tau \cdot {\pi}$, $\rho =
{\tau} \cdot {\rho}$. The definitions of the charmed meson
iso-doublets are
\begin{eqnarray}
\bar{D}^T &=& \left( \bar{D}^0 ,  D^- \right), \qquad D = \left(
D^0 , D^+ \right),
\nonumber \\
 \bar{D}^{*T} &=& \left( \bar{D}^{*0} ,  D^{*-} \right), \qquad D^*
= \left( D^{*0} , D^{*+} \right).
\end{eqnarray}
We will present the values of the necessary coupling constants in
the section for the numerical computations.

With the above preparation, we can write out the decay amplitude
involving contributions from the diagrams of Fig. \ref{diagrams}
in terms of the Cutkosky Cutting Rules. For the process $B^0\to
D^-(p_1)\pi^{+}(p_2)\to J/\psi(p_3,\epsilon_3)\bar D^0(p_4)$ where
as shown in Fig. \ref{diagrams} the vector-meson $D^{*+}$ is
exchanged at t-channel, the resultant amplitude is
\begin{eqnarray}
&&{Abs}^{(a)}\nonumber\\&&=\frac{1}{2}\int\frac{d^{3}p_1}{(2\pi)^3
2E_1}\frac{d^{3}p_2}{(2\pi)^3 2E_2}(2\pi)^4
\delta^4(p_{B}-p_1-p_2) \nonumber\\&&\quad\times A( B^0\to
D^-\pi^+) (-i)g_{\psi D^*
D}\varepsilon^{\mu\nu\alpha\beta}(-i)p_{3\mu}\epsilon_{3\nu}\nonumber\\&&\quad\times
[-i(p_3-p_1)_{\alpha}]i^2\sqrt{2}g_{D^*D\pi}
(ip_2^{\xi})\nonumber\\&&\quad\times
\Big(-g_{\xi\beta}+\frac{q_{\xi}q_{\beta}}{m_{D^*}^2}\Big)\frac{i}{q^2-m_{D^*}^2}\mathcal{F}^2[q^2,m^2_{D^*}].
\end{eqnarray}

The amplitudes of the mode $B^0\to
D^-(p_1)\rho^{+}(p_2,\epsilon_{\rho})\to J/\psi(p_3,\epsilon_3)\bar
D^0(p_4)$ where $D^+$ and $D^{*+}$ are exchanged respectively, read
as
\begin{eqnarray}
&&{Abs}^{(b-1)}\nonumber\\&&=\frac{1}{2}\int\frac{d^{3}p_1}{(2\pi)^3
2E_1}\frac{d^{3}p_2}{(2\pi)^3 2E_2}(2\pi)^4
\delta^4(p_{B}-p_1-p_2)\nonumber\\&&\quad\times A( B^0\to
D^-\rho^+) i^2 g_{\psi
DD}\epsilon_{3\mu}[i(p_3-p_1)^{\mu}-ip_1^{\mu}]\nonumber\\&&\quad\times
i^2\sqrt{2}g_{DD\rho}%\underline{\epsilon_{\rho}^{\nu}}
[(-i)p_{4\nu}-(-i)(p_3-p_1)_{\nu}]\nonumber\\&&\quad\times\Big(-g^{\nu\xi}+\frac{p_{2}^{\nu}p_2^{\xi}}{m^2_{\rho}}\Big)
\frac{i}{q^2-m_{D}^2}\mathcal{F}^2[q^2,m^2_{D}]
\end{eqnarray}
and
\begin{eqnarray}
&&{Abs}^{(b-2)}\nonumber\\&&=\frac{1}{2}\int\frac{d^{3}p_1}{(2\pi)^3
2E_1}\frac{d^{3}p_2}{(2\pi)^3 2E_2}(2\pi)^4
\delta^4(p_{B}-p_1-p_2)\nonumber\\&&\quad\times A( B^0\to
D^-\rho^+) (-i)g_{\psi D^*
D}\varepsilon^{\mu\nu\alpha\beta}(-i)p_{3\mu}\epsilon_{3\nu}\nonumber\\&&\quad\times
[-i(p_3-p_1)_{\alpha}](-i)\sqrt{2}g_{D^*D\rho}
\varepsilon^{\sigma\xi\tau\chi}\nonumber\\&&\quad\times
(-i)(p_3-p_1)_{\sigma}%\underline{\epsilon_{D^{*+}\xi}}
(i)p_{2\tau}%\underline{\epsilon_{\rho\chi}}
\Big(-g_{\xi\beta}+\frac{q_{\xi}q_{\beta}}{m_{D^*}^2}\Big)\nonumber\\&&\quad\times
\Big(-g_{\chi\lambda}+\frac{p_{2\chi}p_{2\lambda}}{m_{D^*}^2}\Big)
\frac{i}{q^2-m_{D^*}^2}\mathcal{F}^2[q^2,m^2_{D^*}].
\end{eqnarray}

For Fig. \ref{diagrams} (c), the amplitude of $B^0\to
D^{*-}(p_1,\epsilon_{D^{*+}})\pi^{+}(p_2)\to
J/\psi(p_3,\epsilon_3)\bar D^0(p_4)$ where $D^{*+}$ is exchanged at
t-channel, is
\begin{eqnarray}
&&{Abs}^{(c)}\nonumber\\&&=\frac{1}{2}\int\frac{d^{3}p_1}{(2\pi)^3
2E_1}\frac{d^{3}p_2}{(2\pi)^3 2E_2}(2\pi)^4
\delta^4(p_{B}-p_1-p_2) \nonumber\\&&\quad\times A( B^0\to
D^{*-}\pi^+) (-i^2)g_{\psi D^*D^*}[i(p_3-2p_1)\cdot \epsilon_3
g_{\mu\nu}\nonumber\\&&\quad\times-i(2p_3-p_1)_{\nu}\epsilon_{3\mu}+i(p_1+p_3)_{\mu}\epsilon_{3\nu}]
%\underline{\epsilon_{D^{*+}}^{\mu}\epsilon_{D^{*-}}^{\nu}}
\nonumber\\&&\quad\times
i^2\sqrt{2}g_{D^*D\pi}%\underline{\epsilon_{D^{*+}}^{\alpha}}
(ip_{2\alpha})
\Big(-g^{\alpha\nu}+\frac{p_{1}^{\alpha}p_1^{\nu}}{m^2_{D^*}}\Big)\nonumber\\&&\quad\times
\Big(-g^{\mu\lambda}+\frac{q^{\mu}q^{\lambda}}{m_{D^*}^2}\Big)
\frac{i}{q^2-m_{D^*}^2}\mathcal{F}^2[q^2,m^2_{D^*}].
\end{eqnarray}

The amplitudes of  $B^0\to
D^{*-}(p_1,\epsilon_{D^{*+}})\rho^{+}(p_2,\epsilon_{\rho})\to
J/\psi(p_3,\epsilon_3)\bar D^0(p_4)$ where $D^+$ and $D^{*+}$ are
exchanged respectively are
\begin{eqnarray}
&&{Abs}^{(d-1)}\nonumber\\&&=\frac{1}{2}\int\frac{d^{3}p_1}{(2\pi)^3
2E_1}\frac{d^{3}p_2}{(2\pi)^3 2E_2}(2\pi)^4
\delta^4(p_{B}-p_1-p_2) \nonumber\\&&\quad\times A( B^0\to
D^{*-}\rho^+)(-i)g_{\psi D^*
D}\varepsilon^{\lambda\tau\sigma\gamma}(-i)p_{3\lambda}\epsilon_{3\tau}\nonumber\\&&\quad\times
(i)p_{1\sigma}%\underline{\epsilon_{D^{*-}\gamma}}
(i^2)\sqrt{2}g_{DD\rho}%\underline{\epsilon^{\xi}_{\rho}}
[(-i)p_{4\xi}-(-i)(p_3-p_1)_{\xi}]
\nonumber\\&&\quad\times\Big(-g^{\mu\xi}+\frac{p_{2}^{\mu}p_2^{\xi}}{m^2_{\rho}}\Big)
\Big(-g_{\gamma}^{\nu}+\frac{p_{1\gamma}p_{1}^{\nu}}{m_{D^*}^2}\Big)\nonumber\\&&\quad\times
\frac{i}{q^2-m_{D}^2}\mathcal{F}^2[q^2,m^2_{D}]
\end{eqnarray}
and
\begin{eqnarray}
&&{Abs}^{(d-2)}\nonumber\\&&=\frac{1}{2}\int\frac{d^{3}p_1}{(2\pi)^3
2E_1}\frac{d^{3}p_2}{(2\pi)^3 2E_2}(2\pi)^4
\delta^4(p_{B}-p_1-p_2) \nonumber\\&&\quad\times A( B^0\to
D^{*-}\rho^+) (-i^2)g_{\psi D^*D^*}[i(p_3-2p_1)\cdot \epsilon_3
g_{\lambda\tau}\nonumber\\&&\quad\times-i(2p_3-p_1)_{\tau}\epsilon_{3\lambda}+i(p_1+p_3)_{\lambda}\epsilon_{3\tau}]
%\underline{\epsilon_{D^{*+}}^{\lambda}\epsilon_{D^{*-}}^{\tau}}
\nonumber\\&&\quad\times
(-i)\sqrt{2}g_{D^*D\rho}\varepsilon^{\sigma\xi\omega\chi}(-i)(p_3-p_1)_{\sigma}%\underline{\epsilon_{D^*+\xi}}
(i)p_{2\omega}%\underline{\epsilon_{\rho\chi}}
\nonumber\\&&\quad\times\Big(g^{\nu\tau}+\frac{p_1^{\nu}p_{1}^{\tau}}{m_{D^*}^2}\Big)
\Big(-g^{\mu}_{\chi}+\frac{p_2^{\mu}p_{2\chi}}{m_{\rho}^2}\Big)\nonumber\\&&\quad\times
\Big(-g_{\xi}^{\lambda}+\frac{q_{\xi}q^{\lambda}}{m^2_{D^*}}\Big)\frac{i}{q^2-m_{D^*}^2}\mathcal{F}^2[q^2,m^2_{D^*}].
\end{eqnarray}
In the above expressions, $\mathcal{F}[q^2,m_i^2]$ denotes the
form factor (FF), which reflects the structure effect at the
effective interaction vertices. In this work, following Ref.
\cite{HY-Chen} we take the monopole form for FF  as
\begin{equation}
\mathcal{F}[q^2,m_i^2]=\Big(\frac{\Lambda^2-m_i^2}{\Lambda^2-q^2}\Big),
\end{equation}
where the phenomenological parameter $\Lambda$ can be parameterized
as
\begin{equation}\label{parameter}
\Lambda=m_i^2+\alpha\Lambda_{QCD}. \end{equation} $m_i$ stands as
the mass of the exchanged meson at t-channel which is depicted in
Fig. \ref{diagrams}.

The decay amplitude of $B^0\to J/\psi \bar{D}^0$ via the hadronic
loop diagrams is
\begin{eqnarray}
&&\mathcal{M}[B^0\to D^{(*)-}\pi^+(\rho^+)\to J/\psi
\bar{D}^0]\nonumber\\&&=Abs^{(a)}+Abs^{(b-1)}+Abs^{(b-2)}\nonumber\\&&\quad+Abs^{(c)}+Abs^{(d-1)}+Abs^{(d-2)}.
\end{eqnarray}

%%%%%%%%%%%%%%%%%%%%%%%%%%%%%%%%%%%%%%%%%%%%%%%%%%%%%%%%%%%%%%%%%%%%%%%
\section{ $B\to D^{(*)}\pi(\rho)$ decays in the factorization
approach}\label{sec2}
%%%%%%%%%%%%%%%%%%%%%%%%%%%%%%%%%%%%%%%%%%%%%%%%%%%%%%%%%%%%%%%%%%%%%%%%

In this section, let us turn to study  $B\to D^{(*)}\pi(\rho)$. We
can reliably apply the factorization  which  has been proved to be
valid to all orders of the strong coupling constant in the heavy
quark limit \cite{BBNS,SCET}, to calculate the amplitude. The
essential non-perturbative quantities are light meson ($\pi,\rho$)
decay constants and $B\to D(D^*)$ transition form factors.

The decay constants for pseudoscalar ($P$) and vector ($V$) mesons
are defined as
 \beq \label{dc}
 \la P(q)|A_{\mu}|0\ra&=&-if_P q_{\mu}, \non\\
 \la V(q,\epsilon)|V_{\mu}|0\ra&=&f_Vm_V\epsilon^*_{\mu}.
 \eeq
where vector and axial vector currents are $V_{\mu}=\bar
q_1\gamma_{\mu}q_2$ and  $A_{\mu}=\bar q_1\gamma_{\mu}\gamma_5q_2$
respectively  and $\epsilon$  is the polarization vector of $V$.

The $B\rightarrow H\left( H=D,\ D^{*}\right)$ transition form
factors are conventionally parameterized as in \cite{CGW}
\begin{eqnarray}
&&\langle D| V_{\mu} | \bar{B} \rangle=
F_{1}(q^2)\Big\{P_{\mu}-\frac{P\cdot q }{q^2}q_{\mu} \Big\}
\nonumber\\&&\qquad\quad\qquad+\frac{P\cdot q}{q^2}F_{0}(q^2)\,q_{\mu}, \label{ffp} \\
%%%%%%%%%%%%%%%%%%%%%%%%%%%%%%%%%%%
&&\langle D^{*}(\epsilon )| V_{\mu} | \bar{B}%
\rangle=\frac{V(q^{2})}{m_{B}+m_{D^{*}}}\varepsilon _{\mu
\alpha \beta \rho }\epsilon ^{*\alpha }P^{\beta }q^{\rho },  \\
&&\langle D^{*}(\epsilon )| A_{\mu} | \bar{B}
\rangle =i\Big[2m_{D^{*}}A_{0}(q^{2})\frac{\epsilon ^{*}\cdot q}{%
q^{2}}q_{\mu }\nonumber\\&&\qquad\quad+( m_{B}+m_{D^{*}})
A_{1}(q^{2})\Big( \epsilon
_{\mu }^{*}-\frac{\epsilon ^{*}\cdot q}{q^{2}}q_{\mu }\Big)  . \nonumber \\
&&\qquad\quad-A_{2}(q^{2})\frac{\epsilon ^{*}\cdot q}{m_{B}+m_{D^{*}}}\Big( P_{\mu }-%
\frac{P\cdot q}{q^{2}}q_{\mu }\Big)\Big],  \label{ffv}
\end{eqnarray}
where $P=p_{B}+p_{D^{(*)}}$, $q=p_{B}-p_{D^{(*)}}$ and $P \cdot
q=m^{2}_{B}-m^{2}_{D^{(*)}}$.

 The decay amplitudes for $B\to
D^{(*)}\pi(\rho)$ are given as \beq
 &&A( B^0\to D^-\pi^+)\nonumber\\&&=\frac{G_F}{\sqrt 2}V_{cb}^*V_{ud}a_1
 \left\{if_\pi(m^2_{B}-m^2_D)F^{BD}_{0}(m^2_\pi)\right\},\label{1}
\eeq

\beq && A( B^0\to D^-\rho^+)\nonumber\\&&=\frac{G_F}{\sqrt
2}V_{cb}^*V_{ud}a_1
 \left\{2f_\rho\,m_\rho F_1^{BD}(m_{\rho}^2)(\varepsilon_\rho^*\cdot p_{_{B}})
 \right\},
\eeq

\beq && A( B^0\to D^{*-}\pi^+)\nonumber\\&&=\frac{G_F}{\sqrt
2}V_{cb}^*V_{ud}a_1
 \left\{ 2f_\pi\,m_{D^*} A_0^{BD^*}(m_{\pi}^2)(\varepsilon_{D^*}^*\cdot p_{_{B}})
 \right\},\nonumber\\
\eeq

\beq
 &&A(B^0\to D^{*-}\rho^+)\nonumber\\&&=\frac{G_F}{\sqrt 2}V_{cb}^*V_{ud}a_1
\Bigg\{- if_{{\rho}}m_{{\rho}}\Bigg[
(\varepsilon^*_{D^*}\cdot\varepsilon^*_{\rho})
 (m_{B}\nonumber \\&&\quad+m_{D^*})A_1^{ BD^*}(m_{\rho}^2)  \nonumber \\&&
\quad- (\epsilon^*_{D^*}\cdot p_{\rho})(\epsilon^*_{\rho} \cdot
p_{_{D^*}})
 {2A_2^{BD^*}(m_{\rho}^2)\over (m_{B}+m_{D^*})}
\nonumber \\&&\quad
+i\epsilon_{\mu\nu\alpha\beta}\varepsilon^{*\mu}_{\rho}
 \varepsilon^{*\nu}_{D^*}p^\alpha_{\rho}
 p^\beta_{D^*}\,{2V^{B{D^*}}(m_{\rho}^2)\over (m_{B}+m_{D^*})
 }\Bigg]\Bigg\}.\label{4}
\eeq

%%%%%%%%%%%%%%%%%%%%%%%%%%%%%%%%%%%%%%%%%%
\section{Numerical result}\label{sec4}
%%%%%%%%%%%%%%%%%%%%%%%%%%%%%%%%%%%%%%%%%%

The relevant input parameters which are employed in this work
include: $m_{B^0}=5279.4$ MeV, $m_{D^0}=1864.5$ MeV,
$m_{J/\psi}=3096.9$ MeV, $m_{D^{\pm}}=1869.3$ MeV,
$m_{D^{*\pm}}=2010.0$ MeV, $m_{\pi^{\pm}}=139.6$ MeV,
$m_{\rho^{\pm}}=775.5$ MeV, $G_{F}=1.16637\times 10^{-5}$
GeV$^{-1}$, $V_{ud}=0.974$, $V_{cb}=41.6\times 10^{-3}$ \cite{PDG};
$g_{D^*D\pi}=8.84$, $g_{D^*D^*\pi}=9.08$ GeV$^{-1}$, $g_{\psi
DD}=g_{\psi D^*D^*}=7.71$, $g_{\psi D^*D}=8.64$,
$g_{D^*D^*\rho}=g_{DD\rho}=2.52$, $g_{D^*D\rho}=2.82$ GeV$^{-1}$
\cite{Lagarangian}; $f_{\pi}=132$ MeV and $f_{\rho}=216$ MeV
\cite{CCH}.

The Wilson coefficient $a_1$  has been calculated up to the
next-to-leading order \cite{BBNS} and we take the value \beq
 a_1=1.05.
\eeq

The momentum dependence of the transition form factors in eqs.
(\ref{1})-(\ref{4}) possess the pole structures \cite{CCH} as
\begin{eqnarray}
F(q^2)=\frac{F(0)}{1-a\zeta+b\zeta^2}
\end{eqnarray}
with $\zeta=q^2/m_{B}^{2}$.  $F(0)$, $a$ and $b$ are obtained by
fitting data and their values are shown in Table \ref{table-1}.
\begin{table}[htb]
\begin{center}
 \begin{ruledtabular}
\begin{tabular}{c|ccc}
$F$&$F(0)$&$a$&$b$\\\hline $F_{0}^{BD}$&0.67&0.65&0.00\\
$F_{1}^{BD}$&0.67&1.25&0.39\\
$A_{0}^{BD^*}$&0.64&1.30&0.31\\
$A_{1}^{BD^*}$&0.63&0.65&0.02\\
$A_{2}^{BD^*}$&0.61&1.14&0.52\\
$V^{BD^*}$&0.75&1.29&0.45\\
\end{tabular}
 \end{ruledtabular}
\end{center}\caption{The values of $F(0)$, $a$ and $b$ in the form factors of
$B\to D^{(*)}$ \cite{CCH}. \label{table-1}}
\end{table}
With the values given in Table \ref{table-1}, one obtains
$F_{0}^{BD}(m^2_{\pi})=0.67$, $F_{1}^{BD}(m^2_{\rho})=0.69$,
$A_{0}^{BD^*}(m^2_{\pi})=0.64$, $A_{1}^{BD^*}(m^2_{\rho})=0.64$,
$A_{2}^{BD^*}(m^2_{\rho})=0.63$, $V^{BD^*}(m^2_{\rho})=0.77$, which
will be applied to the later numerical calculation.

\begin{figure}[htb]
\begin{center}
\begin{tabular}{c}
\scalebox{0.85}{\includegraphics{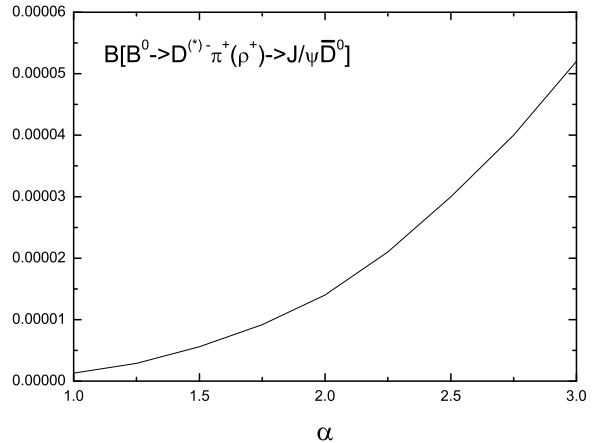}}\\
\end{tabular}
\end{center}
\caption{The variation of the branching ratio of $B^0\to
J/\psi\bar{D}^0$ with $\alpha=1\sim 3$. \label{DP}}
\end{figure}

In Fig. \ref{DP}, we show the dependence of the branching ratio of
$B^0\to J/\psi \bar{D}^0$ on the phenomenological parameter $\alpha$
in eq. \ref{parameter}, which spans a range $\alpha=1\sim 3$.
Furthermore, Table \ref{TP} presents the branching ratio of $B^0\to
J/\psi \bar{D}^0$ with some typical values of $\alpha$.

\begin{table}[htb]
\begin{center}
 \begin{ruledtabular}
\begin{tabular}{c|ccccc}
$\alpha$&1.0&1.5&2.0&2.5&3.0\\\hline
$B[B^0\to J/\psi \bar{D}^0]$&&&&&\\
$(\times10^{-5})$&$0.13$&$0.54$&$1.4$&$3.0$&$5.2$
\end{tabular}
 \end{ruledtabular}
\end{center}\caption{The branching ratio of $B^0\to J/\psi \bar{D}^0$ corresponding to several
typical values of $\alpha$. \label{TP}}
\end{table}

\section{Conclusion and discussion}

In this work, we calculate the contribution of the FSI, i.e.
inelastic rescattering processes to the branching ratio of $B^0\to
J/\psi \bar{D}^0$ and find that it spans a relatively wider range
of $10^{-6}\sim 10^{-5}$ which is obviously larger than the
theoretically predicted value $10^{-8}$ \cite{EMY} and  comparable
with the pQCD prediction of $10^{-6}$ \cite{LLQ}, moreover, it is
also noted that as $\alpha=3$ is taken, it can be close to the
experimental upper bounds. We hope that the Babar and Belle
collaborations will further investigate the process and carry out
more accurate measurements. Then not only an upper bounds will be
given, but also a rather precise branching ratio can be obtained.

The significance of this investigation to the present theoretical
frameworks is obvious as discussed in the section of introduction.
Since in the process the linear momentum of the final products is
not large, there can exist a region where the exchanged gluon is
soft and application of pQCD might fail. This region should be fully
governed by the non-perturbative QCD effects which are not involved
in the conventional pQCD calculations, even though phenomenological
wavefunctions of the hadrons can partly cover such effects. Thus we
consider the FSI effects as an additional contribution to that of
pQCD evaluation. However, on  another aspect, one cannot indeed
determine the range where pQCD fails and this is exactly the goal of
this work.

As noticed, there exist some phenomenological parameters in our
calculations on FSI effects, such as $\alpha$ or $\Lambda$ in eq.
(\ref{parameter}) and other uncertainties which are coming from
the employed data, therefore, we can only trust the results to the
order of magnitude. However, the largeness of the contribution of
the FSI should draw our attention because it may change the whole
scenario. In fact, an accurate measurement can provide an ideal
field for testing validity of pQCD. Since the FSI can result in a
branching ratio as large as $10^{-6}\sim 10^{-5}$, there is a
region where the pQCD prediction and the contribution of the FSI
have the same order, thus it is hard to clearly identify
individual contribution from both mechanism, unless accurate
measured data are available. On other aspect, under the assumption
that the present pQCD calculation is trustworthy to a certain
accuracy, it is not hopeless to determine their fractions because
the two factories indeed have ability to carry out such precise,
but difficult measurements.

More concretely, if the future measurement confirmed a smaller
branching ratio of about $10^{-6}$, then one would make a careful
analysis to distinguish between the two kinds of contributions.
Furthermore, if the data are basically consistent with the pQCD
predicted value, it is indicated that pQCD works well, even though
there might be a range where application of pQCD is dubious. In
other words, for that case, the range where pQCD fails, does not
make dominant contribution and the whole theoretical framework
should cover a much wider application range than was expected.
Then we have to re-adjust the input parameters for calculating the
FSI or set a more stringent constraint on them. It would be very
helpful for gaining knowledge on the FSI which plays important
roles in many decay and production processes.

By contraries, if the future measurements on $B^0\to J/\psi
\bar{D}^0$ confirm that the branching ratio is obviously larger
than $10^{-6}$, the fact would indicate that the region where pQCD
fails, is important and should be re-considered. In that case, we
can conclude that one must be careful as he applies the pQCD to
evaluate physical processes with low energy scales. And the FSI
may be a possible solution to the discrepancy. If it is true, a
byproduct would be that one can further investigate details about
the methods for calculation on the FSI effects and determine the
concerned parameters, since the "contamination" from the direct
process which is evaluated in terms of pQCD is relatively small.

As a conclusion, we would urge our experimental colleagues to make
a more accurate measurement on this process because its
significance to our theory is obvious.

%%%%%%%%%%%%%%%%%%%%%%%%%%%%%%%%
\section*{Acknowledgments}
%%%%%%%%%%%%%%%%%%%%%%%%%%%%%%%%

This work is partly supported by the National Natural Science
Foundation of China (NNSF) and a special grant of the Education
Ministry of China. X.L. was also supported by the
\emph{Funda\c{c}\~{a}o para a Ci\^{e}ncia e a Tecnologia of the
Minist\'{e}rio da Ci\^{e}ncia, Tecnologia e Ensino Superior} of
Portugal
(SFRH/BPD/34819/2007).\\

%\end{CJK}
\end{document}